\documentclass{article}
\usepackage{amsmath}
\usepackage{hyperref}
\begin{document}
\title{The slowly passage through the resonances and wave packets with the different carriers\thanks{This work was supported by grants RFBR 03-01-00716,
Leading Scientific Schools 1446.2003.1 and INTAS 03-51-4286.}}
\author{Oleg Kiselev\thanks{Institute of Math. USC
RAS; ok@ufanet.ru}, Sergei Glebov\thanks{Ufa State Petroleum Technical University; sg@anrb.ru}}

\maketitle

\begin{abstract}
Solution of the nonlinear Klein-Gordon equation perturbed by small
external force is investigated. The perturbation is represented by finite collections of harmonics. The frequencies of the perturbation vary
slowly and pass through the resonant values consecutively. The resonances  lead to the sequence  of the wave packets with the different fast oscillated carriers. Full asymptotic description of this process is presented.
\end{abstract}

\font\Sets=msbm10
\def\Real{\hbox{\Sets R}}
\def\Complex{\hbox{\Sets C}}
\def\bb{\begin{equation}}
\def\ee{\end{equation}}
\def\pt{\partial}
\def\mod{\hbox{mod}}
\def\const{\hbox{const}}
\def\sgn{\hbox{sgn}}
\def\Arg{\hbox{Arg}}
\def\a{\alpha}
\def\b{\beta}
\def\d{\delta}
\def\G{\Gamma}
\def\g{\gamma}
\def\e{\epsilon}
\def\ve{\varepsilon}
\def\k{\kappa}
\def\l{\lambda}
\def\O{\Omega}
\def\o{\omega}
\def\th{t}
\def\s{\sigma}
\def\t{\tau}
\def\z{\zeta}
\newtheorem{lemma}{\bf Lemma}
\newtheorem{theorema}{\bf Theorem}

\noindent{\bf Introduction}
\par
In this work we study the problem of a generation of sequences of solitary packets with different carriers in the optical fiber. The nonlinear Klein-Gordon equation is studied as a modeling equation. We consider this equation which is perturbed by a small external driving force with finite collection of modes.  The wave packets appear due to passage through a resonance by different modes of the external force.  After the passage through the whole resonances the solution contains the full collection of the solitary packages of waves with different carriers. The envelope functions of these packages satisfy to nonlinear Schr\"odinger equation (NLSE). 
\par
In general the derivation of NLSE for small solutions of nonlinear equations is well known \cite{Kelley,Talanov,Zakharov} and  justified \cite{Kalyakin0}. Our solution has a more complicate structure. Before all resonances the solution has an order of the perturbation and defined by the external force. After the passage through the resonance of the last mode of the perturbation the solution has the order of the square root of the order of the perturbation and satisfies NLSE. So we show the process of the resonant transformation of the solution and the appearance of the wave packets with the different carriers. 
\par
Earlier the resonant generation of periodic waves  by a small external force was investigated by a computer simulation \cite{Friedland-Shagalov}.  The phenomenon of the generation and the scattering of the solitary waves in the case of nonlinear Schrodinger equation was asymptotically investigated in \cite{Glebov-Kiselev-Lazarev,GK}. The problem on the generation of a solitary wave with a single carrier was solved in \cite{GKL}. Here we use the proposed approach to study the multiphase case. 
\par
The proposed approach is based on  a local resonance phenomenon. The
local resonance in linear ordinary differential equations was
investigated in papers \cite{Kevorkyan,Rozenfeld}. Later this
phenomenon was investigated in partial differential equations for
linear case \cite{Neu} and for weak nonlinear case
\cite{Kalyakin1,Glebov}. 
\par
The goal of this paper is the following: to demonstrate that the passage through the resonances allows to obtain the sequences of solitary packets of waves with the different carriers. The packets with different carriers do not interact. 
\par
This paper has the following structure. The first section contains
the statement of the problem, main result and example. The second section contains the
asymptotic construction before the resonance. In the third
section we construct an internal asymptotics in the neighborhood of
the resonance curve. In the fourth section we construct the asymptotic expansion after the passage through the resonance. Five section contains the analysis of the resonances in high-order terms of the asymptotic solution. All asymptotics are matched.
\par

\section{Main result}

\par
Let us consider the Klein-Gordon equation with the cubic nonlinearity
perturbed by external force with the finite collection of the different harmonics
\bb
\pt^2_{t} U - \pt^2_{x} U + U + \gamma U^3 = \ve^2
\sum_{k=1}^{N}f_k(\ve x)\exp\left\{i k {{S(\ve^2t,\ve^2x)}\over
{\ve^2}}\right\}\,+c.c.,\quad 0 < \ve \ll 1. \label{we}
\ee
Here $\gamma=\const$;  $f_k(y),\ k=1,\dots,N$ and the phase function $S(y,z)$ are smooth.
\par
We construct the formal asymptotic solution of the WKB-type.
In the domain when the first mode of the perturbation does not pass through the resonance the asymptotic solution has a form
\bb
U= - \ve^2 \sum_{j=1}^N {{f_j}\over{l_j}}\exp\{i (jS)/\ve^2\} + O(\ve^3), \label{shortForcedOscilations}
\ee
where
$$
l_j(x_2,t_2) \equiv j^2 (\pt_{t_2}S)^2 - j^2 (\pt_{x_2}S)^2 -1,\quad x_2=\ve^2x,\ t_2=\ve^2t.
$$
The leading-order term has an order of $\ve^2$ and oscillates. Such solution relates to the  forced oscillations. 
\par
On the curve $l_k(x_2,t_2)=0,\ k=N,\dots,1$ the  frequencies of $k$-th mode of the forced oscillations and a frequency of the eigen oscillations of the linearized Klein-Gordon equation are equal. Usually the curves with the such property are called resonant curves. The local resonance layer appears  in the neighborhood of the curve $l_k(x_2,t_2)=0$. Here we describe the passage through the resonant layer with the ordinary number $k$.
\par
After passage through the $k$-th resonant layer a new eigen mode with the amplitude of the order $\ve$ appears in the solution. It leads to changing of the number of harmonics in leading-order term. The amplitudes of the leading-order term satisfy $(N-k+1)$ nonlinear Schr\"odinger equations after passage through the $k$-th resonance. 
The solution has a form
\bb
U=\ve \sum_{j=k}^N \stackrel{1,0}{\Psi}{}_{\Phi_j} \exp\{i\Phi_{j}/\ve^2\} + O(\ve^2), \label{shortanzats}
\ee
here the value $k$ in (\ref{shortanzats}) is defined by
$$
k= \min j : l_j >0,\quad j=1,\dots,N 
$$
and  functions $l_j(x_2,t_2)$ are arrayed: 
$$
l_j(x_2,t_2)<l_m(x_2,t_2), \quad 1 \le j < m \le N.
$$
The accurate formulation of the result for this paper is following
\par
\begin{theorema}\label{mainTheoremInTermsOfPhases}
Let the asymptotic solution of (\ref{we}) relates to forced oscillations:
$$
U= - \ve^2 \sum_{j=1}^N {{f_j}\over{l_j}}\exp\{i (jS)/\ve^2\} + O(\ve^3)
$$
in the domain $l_N <0$ before the passage through the first resonance 
and the asymptotic solution has a form 
$$
U=\ve \sum_{j=k+1}^N \stackrel{1,0}{\Psi}{}_{\Phi_j} \exp\{i\Phi_{j}/\ve^2\} + O(\ve^2),
$$
in the domain $-l_k \gg \ve,\quad l_{k+1}\gg \ve$ 
before the $k$-th resonance curve $l_k=0$.
Here 
$$
\Phi_{j} \in \Upsilon_{1,0}^{b}=\{\pm\Phi_{k+1},\pm\Phi_{k+2},\dots,\pm\Phi_{N}\}
$$ 
and the amplitudes $\stackrel{1,0}{\Psi}{}_{\Phi_j}$ satisfies  $(N - k)$ nonlinear Schrodinger equations
\begin{eqnarray}
2i\pt_{t_2}\Phi_j\pt_{t_2}\stackrel{1,0}{\Psi}{}_{\Phi_j} +
\pt^2_{\xi_j}\stackrel{1,0}{\Psi}{}_{\Phi_j} +  i[\pt_{t_2}^2\Phi_j -
\pt_{x_2}^2\Phi_j]\stackrel{1,0}{\Psi}{}_{\Phi_j} + \label{systemNLSEbefore} \\ 
\gamma |\stackrel{1,0}{\Psi}{}_{\Phi_j}|^2 \stackrel{1,0}{\Psi}{}_{\Phi_j}
=0,\quad j=k+1,\dots,N. \nonumber
\end{eqnarray}
Then in the domain $-l_{k-1} \gg \ve,\quad l_{k}\gg \ve$ the solution has the similar structure with the changes of the phase collection according to
$$
\Upsilon_{1,0}^{p}=\Upsilon_{1,0}^{b}\cup \{\pm \Phi_k \}.
$$
The number of equations in (\ref{systemNLSEbefore}) increases up to $(N - k +1)$.
\end{theorema}
Here we use the upper indexes ${}^b$ and ${}^p$ in our notations to specify the items in the domain {\bf before} the passage through the $k$-th resonant curve and {\bf post} resonance items.
\par
To illustrate the the main result let us consider equation (\ref{we}) with the simplest driving force with two modes:
$$
F=f_1(x_1)\exp(i(t_2^2/2+x_2)/\ve^2)+f_2(x_1)\exp(i(t_2^2+2x_2)/\ve^2).
$$
In this case the curves of the local resonance are the lines $t_2=\sqrt{17/16}$ and $t_2=\sqrt{2}$. 
In the domain $t_2<\sqrt{17/16}$ the asymptotic solution is
$$
U = O(\ve^2).
$$
The solution of this order with respect to $\ve$ relates to forced oscillations.
\par
After the passage through the first resonance  the solution is
$$
U = \ve \stackrel{1,0}{\Psi}{}_{\Phi_2} \exp\{i \Phi_2 /\ve^2 \} + O(\ve^2),\quad \sqrt{17/16}<t_2<\sqrt{2}
$$
The amplitude $\stackrel{1,0}{\Psi}{}_{\Phi_2}$ satisfies the Cauchy problem for
$$
2i\pt_{t_2}\stackrel{1,0}{\Psi}{}_{\Phi_2} +
\pt^2_{\xi_1\xi_1}\stackrel{1,0}{\Psi}{}_{\Phi_2} + \gamma
|\stackrel{1,0}{\Psi}{}_{\Phi_2}|^2 \stackrel{1,0}{\Psi}{}_{\Phi_2}
=0,
$$
variable $\xi_1=t_1-\sqrt{17/16}x_1$, Initial condition is
$$
\stackrel{1,0}{\Psi}{}_{\Phi_2}|_{t_2=\sqrt{17/16}}= f_1(\xi)(1+i)\sqrt{\pi}.
$$
After the passage through the second resonance the solution is 
$$
U =\ve \stackrel{1,0}{\Psi}{}_{\Phi_1} \exp\{i \Phi_1 /\ve^2 \} + \ve \stackrel{1,0}{\Psi}{}_{\Phi_2} \exp\{i \Phi_2 /\ve^2 \} + O(\ve^2),\quad t_2 > \sqrt{2}.
$$
The amplitudes  $\stackrel{1,0}{\Psi}{}_{\Phi_1}$ and $\stackrel{1,0}{\Psi}{}_{\Phi_2}$ are determined from  the Cauchy problem for two  nonlinear Shr\"odinger equation:
$$
2i\pt_{t_2}\stackrel{1,0}{\Psi}{}_{\Phi_1} +
\pt^2_{\xi_1\xi_1}\stackrel{1,0}{\Psi}{}_{\Phi_1} + \gamma
|\stackrel{1,0}{\Psi}{}_{\Phi_1}|^2  \stackrel{1,0}{\Psi}{}_{\Phi_1}
=0,
$$
$$
2i\pt_{t_2}\stackrel{1,0}{\Psi}{}_{\Phi_2} +
\pt^2_{\xi_2\xi_2}\stackrel{1,0}{\Psi}{}_{\Phi_2} + \gamma
|\stackrel{1,0}{\Psi}{}_{\Phi_2}|^2\stackrel{1,0}{\Psi}{}_{\Phi_2}
=0.
$$
Here $\xi_2=t_1-\sqrt{2}x_1/2$, initial conditions are 
$$
\stackrel{1,0}{\Psi}{}_{\Phi_2}|_{t_2=\sqrt{2}}= \stackrel{1,0}{\Psi}{}_{\Phi_2}|_{t_2=\sqrt{2}-0},\quad
\stackrel{1,0}{\Psi}{}_{\Phi_1}|_{t_2=\sqrt{2}}= f_2(\xi_2)(1+i)\sqrt{\pi}.
$$
The solution of this Cauchy problem contains solitary waves if the
initial data are sufficiently large \cite{TeoriyaSolitonov}.

\section{Pre-resonance expansion}
\label{externalAsymptotics1}
\par
In this section we construct  the formal asymptotic solution  in the
domain before the $k$-th resonant curve. Here we use scaled variables $x_j=\ve^j x, t_j=\ve^j t, j=1,2$.
 The solution has a form 
\bb
U(x,t,\ve)=\ve \Psi^b_{1}+ \sum_{n=2}^\infty\ve^n \Psi^b_{n},\label{solution_before_k}
\ee
$$
\Psi^b_{1}=\sum_{m=k+1}^{N} \sum_{\pm\Phi_m}\exp\{\pm
i\Phi_m(x_2,t_2)/\ve^2\} \stackrel{1,0}{\Psi}{}_{\pm\Phi_m}(x_1,t_1,t_2)
$$
The higher order terms are
\begin{eqnarray*}
\Psi^b_{n} = \sum_{j=0}^{n-2} \ln^j(\ve)\sum_{m=k+1}^{N}
\bigg(\sum_{\pm\Phi_m}\exp\{\pm
i\Phi_m(x_2,t_2)/\ve^2\} \stackrel{n,j}{\Psi}{}_{\pm\Phi_m}(x_1,t_1,t_2)+ \\
 \sum_{\chi \in
\Upsilon^{b}{}'_{n,j}}\exp\{i\chi(x_2,t_2)/\ve^2\}
\stackrel{n,j}{\Psi}{}_{\chi}(x_1,t_1,t_2)\bigg),
\end{eqnarray*}
where $\Upsilon^{b}{}'_{n,j}$ is a set of phase functions which is determined by 
$$
\Upsilon^{b}_{1,0}=\{\pm\Phi_{k+1},\pm\Phi_{k+2},\dots,\pm\Phi_{N}\};\quad \Upsilon^{b}_{2,0}= \Upsilon^{b}_{1,0} \cup \{\pm S,\dots, \pm N S\},
$$
\bb
\Upsilon^{b}_{n,j}=\bigcup_{\begin{array}{c} n_1+n_2+n_3=n,\\ j_1+j_2+j_3=j \end{array}}
\chi_{n_1,j_1}+\chi_{n_2,j_2}+\chi_{n_3,j_3},\,\,\,\chi_{n_p,j_p} \in \Upsilon^{b}_{n_p,j_p}.\nonumber
\ee
$$
\Upsilon^{b}{}'_{n,j} = \Upsilon^{b}_{n,j} \backslash \Upsilon^{b}_{1,0}.
$$
This solution contains the two parts. 
The first part of the solution has the leading-order term of the order of $\ve$. The phase set of this part is $\Upsilon^{b}_{1,0}=\{\pm\Phi_{k+1},\pm\Phi_{k+2},\dots,\pm\Phi_{N}\}$. The modes have appeared in the solution due to the crossing of the previous $(N-k)$ resonant curves $l_j=0$. The terms of the order  of $\ve^2$ with the phases $\notin \Upsilon^{b}_{1,0}$ relate to the forced oscillations and describe the behaviour of the solution before the first resonance curve $l_{N}=0$ where the forced oscillations  take place only.
\par
The asymptotics constructed in this section is valid as $l_{k+1}\gg \ve,\ -l_{k}\gg\ve$. The result of this section  is formulated below.
\par
{\bf Note.} Expansion (\ref{solution_before_k}) contains the terms $\ve^n\ln^m \ve$. It looks a little bit unexpectedly because of the original equation does not contain the logarithmic terms with respect of $\ve$. But these terms naturally appear due to the passage through the previous resonant layers, see Lemma \ref{lemmaAboutAsymptoticsAsLambdaToInfonity} in subsection \ref{asymptoticsOfHOT}.
\par
Let us construct the formal asymptotic solution for equation
(\ref{we}) in form (\ref{solution_before_k}).
We substitute (\ref{solution_before_k}) in equation (\ref{we}) and collect
the terms of the same order of $\ve$. As a result we obtain a
recurrent sequence of equations for the coefficients of the asymptotics.
\par
Terms of the order of $\ve$ give us the equations for the phase functions
\bb
(\pt_{t_2}\Phi_j)^2 - (\pt_{x_2}\Phi_j)^2 -1 =0,\quad j=k+1,\dots,N. \label{equations_for_phases_pre}
\ee
Initial data for the phase functions $\Phi_j,\ j=k+2,\dots,N$ are determined by their values on the curve $l_{k+1}=0$. 
Initial data for $\Phi_{k+1}$ is related with 
\bb
\Phi_j|_{l_{k+1} = 0} = (k+1)S|_{l_{k+1}=0},\quad
\pt_{t_2}\Phi_j|_{l_{k+1}=0}=(k+1)\pt_{t_2}S|_{l_{k+1}=0}. \label{inialDataForPhases}
\ee
The terms of the order of $\ve^2$ give us the homogeneous transport equation
\bb
\pt_{t_2}\Phi_j\pt_{t_1}\stackrel{1,0}{\Psi}{}_{\Phi_j} -
\pt_{x_2}\Phi_j\pt_{x_1}\stackrel{1,0}{\Psi}{}_{\Phi_j} =0.\quad j=k+1,\dots,N.
\label{perenos_pre}
\ee
This equation allows us to determine the dependence of the
leading-order term  on characteristic variable $\zeta$. Equation
(\ref{perenos_pre}) along the characteristics
\bb
{{d x_1}\over{d \zeta_j}} = - \pt_{x_2}\Phi_j, \quad  {{d t_1}\over{d
\zeta_j}} =  \pt_{t_2}\Phi_j
 \label{charactereq_before}
\ee
can be written in the form of ordinary differential equation
\bb
{{d \stackrel{1,0}{\Psi}{}_{\Phi_j}}\over{d \zeta_j}}=0.
\label{eqInFastVariablesForPsij}
\ee
It yields $\stackrel{1,0}{\Psi}{}_{\Phi_j}$ depends on $\xi_j$, where
the $\xi_j$ is defined by
$$
{{dx_1}\over{d\xi_j}}=\pt_{t_2}\Phi_j,\quad
{{dt_1}\over{d\xi_j}}=\pt_{x_2}\Phi_j.
$$
\par
Among the terms of the order $\ve^3$ we collect the terms which oscillate as
$\exp(i\Phi_j/\ve^2)$. It gives 
\begin{eqnarray*}
2i\left(\pt_{t_2}\Phi_j\pt_{t_1}\stackrel{2,0}{\Psi}{}_{\Phi_j} -
\pt_{x_2}\Phi_j\pt_{x_1}\stackrel{2,0}{\Psi}{}_{\Phi_j}\right) + \nonumber\\
2i\pt_{t_2}\Phi_j\pt_{t_2}\stackrel{1,0}{\Psi}{}_{\Phi_j} +
[(\pt_{t_1}\xi_j)^2-(\pt_{x_1}\xi_j)^2]\pt^2_{\xi_j\xi_j}\stackrel{1,0}{\Psi}{}_{\Phi_j}
+ \\
 i[\pt_{t_2}^2\Phi_j - \pt_{x_2}^2\Phi_j]\stackrel{1,0}{\Psi}{}_{\Phi_j} +
\gamma |\stackrel{1,0}{\Psi}{}_{\Phi_j}|^2
\stackrel{1,0}{\Psi}{}_{\Phi_j} =0.
\end{eqnarray*}
It is convenient to write this equation  in the form of ordinary
differential equation  in terms of characteristic variables
\begin{eqnarray}
{{d \stackrel{2,0}{\Psi}{}_{\Phi_j}}\over{d \zeta_j}} =
-2i\pt_{t_2}\Phi_j\pt_{t_2}\stackrel{1,0}{\Psi}{}_{\Phi_j} -
[(\pt_{t_1}\xi_j)^2-(\pt_{x_1}\xi_j)^2]\pt^2_{\xi_j\xi_j}\stackrel{1,0}{\Psi}{}_{\Phi_j}
\nonumber\\
-  i[\pt_{t_2}^2\Phi_j -
\pt_{x_2}^2\Phi_j]\stackrel{1,0}{\Psi}{}_{\Phi_j} - \gamma
|\stackrel{1,0}{\Psi}{}_{\Phi_j}|^2 \stackrel{1,0}{\Psi}{}_{\Phi_j}.
\label{laste_pre}
\end{eqnarray}
Equation (\ref{eqInFastVariablesForPsij}) shows  that the right hand
side of equation (\ref{laste_pre}) does not depend on $\zeta_j$. 
To construct the bounded solution of (\ref{laste_pre}) we  demand  the right hand side of the 
equation is equal to zero. It allows to determine the dependence of
the leading-order term  on slow variable $t_2$. The amplitudes of the leading-order terms satisfy $(N-k)$ nonlinear Schr\"odinger equations
\begin{eqnarray}
2i\pt_{t_2}\Phi_j\pt_{t_2}\stackrel{1,0}{\Psi}{}_{\Phi_j} +
[(\pt_{t_1}\xi_j)^2-(\pt_{x_1}\xi_j)^2]\pt^2_{\xi_j\xi_j}\stackrel{1,0}{\Psi}{}_{\Phi_j}
+  \nonumber\\
+i[\pt_{t_2}^2\Phi_j -
\pt_{x_2}^2\Phi_j]\stackrel{1,0}{\Psi}{}_{\Phi_j} + \gamma
|\stackrel{1,0}{\Psi}{}_{\Phi_j}|^2 \stackrel{1,0}{\Psi}{}_{\Phi_j}
=0,\quad j=k+1,\dots,N \label{nls_pre}
\end{eqnarray}
\par
The initial conditions for $\stackrel{1,0}{\Psi}{}_{\Phi_j}$ are
\begin{eqnarray*}
\stackrel{1,0}{\Psi}{}_{\Phi_j}|_{l_k=0}=\stackrel{1,0}{\Psi}{}_{\Phi_j}|_{l_k=+0}, \,\, j=k+2,\dots,N;\\
\stackrel{1,0}{\Psi}{}_{\Phi_{k+1}}|_{l_{k+1}=0}= \int_{-\infty}^{\infty} d\s
f_{k+1}(x_1)\exp(i\int_0^\s d\mu \l_{k+1}(x_1,t_1,\ve)).
\end{eqnarray*}
Integration in this integral is realized in the line of characteristic direction connected with (\ref{charactereq_before}).
\par
The equations for  the higher-order  terms  are obtained by  the same
manner
$$
2i\left(\pt_{t_2}\Phi_j\pt_{t_1}\stackrel{n+1,k}{\Psi}{}_{\Phi_j} -
\pt_{x_2}\Phi_j\pt_{x_1}\stackrel{n+1,k}{\Psi}{}_{\Phi_j}\right)
=2i\pt_{t_2}\Phi_j\pt_{t_2}\stackrel{n,k}{\Psi}{}_{\Phi_j} -
\pt^2_{\xi_j\xi_j}\stackrel{n,k}{\Psi}{}_{\Phi_j} -
$$
$$
-i[\pt_{t_2}^2\Phi_j -
\pt_{x_2}^2\Phi_j]\stackrel{n,k}{\Psi}{}_{\Phi_j} +
\pt_{t_1}\xi_j\pt^2_{\xi t_2}\stackrel{n-1,k}{\Psi}{}_{\Phi_j}- \gamma
\sum_{k_1,k_2,l_1,l_2,m_1,m_2,\alpha,\beta,\delta}
\stackrel{k_1,k_2}{\Psi}_{\alpha} \stackrel{l_1,l_2}{\Psi}_{\beta}
 \stackrel{m_1,m_2}{\Psi}_{\delta},
$$
where $k_1+l_1+m_1=n+2,\ k_2+l_2+m_2=k,\ \alpha + \beta + \delta =
\Phi_j,\ \alpha\in \Upsilon^{b}_{k_1,k_2},\ \beta\in \Upsilon^b_{l_1,l_2},\
\delta \in \Upsilon^b_{m_1,m_2}.$
\par
To construct the uniform asymptotic expansion with respect to
$\zeta_j$ we obtain the linearized Schrodinger equation for
higher-order  term
\begin{eqnarray}
2i\pt_{t_2}\Phi_j\pt_{t_2}\stackrel{n,k}{\Psi}{}_{\Phi_j} +
\pt^2_{\xi_j\xi_j}\stackrel{n,k}{\Psi}{}_{\Phi_j} +i[\pt_{t_2}^2\Phi_j -
\pt_{x_2}^2\Phi_j]\stackrel{n,k}{\Psi}{}_{\Phi_j} =
\nonumber \\
- \pt_{t_1}\xi\pt^2_{\xi_j t_2}\stackrel{n-1,k}{\Psi}{}_{\Phi_j}
-\gamma \sum_{k_1,k_2,l_1,l_2,m_1,m_2,\alpha,\beta,\delta}
\stackrel{k_1,k_2}{\Psi}_{\alpha} \stackrel{l_1,l_2}{\Psi}_{\beta}
 \stackrel{m_1,m_2}{\Psi}_{\delta},\label{lSh_pre}
\end{eqnarray}
where $k_1+l_1+m_1=n+2,\ k_2+l_2+m_2=k,\ \alpha + \beta + \delta =
\Phi_j,\ \alpha\in \Upsilon^b_{k_1,k_2},\ \beta\in
\Upsilon^b_{l_1,l_2},\ \delta \in \Upsilon^b_{m_1,m_2}.$
\par
The amplitudes $\stackrel{n,k}{\Psi}_{\chi}$ as $\chi \in \Upsilon^{b'}_{n,k}$ 
are determined by algebraic equations
\bb
\left[-(\chi_{t_2})^2 + (\chi_{x_2})^2 + 1
\right]\stackrel{n,k}{\Psi}_{\chi} = \stackrel{n,k}F_{\chi},\quad
\chi\not= \pm \Phi_j, \,\, j=1,\dots,k. \label{algebraic_pre}
\ee
Here the right hand side of the equation depends on previous
terms and their derivatives
$$
\stackrel{n,k}F_{\chi} =
-2i\chi_{t_2}\pt_{t_1}\stackrel{n-1,k}{\Psi}_{\chi} +
2i\chi_{x_2}\pt_{x_1}\stackrel{n-1,k}{\Psi}_{\chi} - 2i \chi_{t_2}
\pt_{t_2}\stackrel{n-2,k}{\Psi}_{\chi} - i\left[\chi_{t_2 t_2} -
\chi_{x_2 x_2} \right]\stackrel{n-2,k}{\Psi}_{\chi}-
$$
\bb
 \pt^2_{t_1
t_2}\stackrel{n-3,k}{\Psi}_{\chi} - \pt^2_{t_2
t_2}\stackrel{n-4,k}{\Psi}_{\chi} -\gamma
\sum_{k_1,k_2,l_1,l_2,m_1,m_2,\alpha,\beta,\delta}
\stackrel{k_1,k_2}{\Psi}_{\alpha} \stackrel{l_1,l_2}{\Psi}_{\beta}
\stackrel{m_1,m_2}{\Psi}_{\delta}, \label{rightHandSideForAlgebraic}
\ee
where $k_1+l_1+m_1=n-4,\ k_2+l_2+m_2=k,\ \alpha + \beta + \delta =
\chi,\ \alpha\in \Upsilon^b_{k_1,k_2},\ \beta\in \Upsilon^b_{l_1,l_2},\
\delta \in \Upsilon^b_{m_1,m_2}.$
\par
{\bf Note.} It's necessary to note that the multiplier $\left[-(\chi_{t_2})^2 + (\chi_{x_2})^2 + 1
\right]$ can be vanished on some curves. It leads to the resonances for the higher order terms of the asymptotics. The passage through the resonances does not change the leading-order terms of the asymptotic solution. We discuss this passage in section \ref{ResonancesInHigherOrders}.  
\par
In this section we pay a special attention on the amplitudes $\stackrel{n,k}{\Psi}_{\chi}$ with the phase function $kS$. These amplitudes have the strongest order singularity on the curve $l_k=0$. The multiplier 
$\left[-(\chi_{t_2})^2 + (\chi_{x_2})^2 + 1\right]$ on the left hand of equation (\ref{algebraic_pre}) equals zero on the curve $l_k=0$. Here we explicitly write out 
a few first correction terms: 
\bb
\stackrel{2,0}{\Psi}_{kS} = - {f_k \over l_k}, \label{2th}
\ee
\bb
\stackrel{3,0}{\Psi}_{kS} = 2ik{{\pt_{x_1}f_k \pt_{x_2}S} \over l_k^2},
\label{3th}
\ee
\begin{eqnarray}
\stackrel{4,0}{\Psi}_{kS}= {{2ikf_k[\pt_{t_2}S \pt_{t_2}l_k -
\pt_{x_2}S\pt_{x_2}l_k] -4k^2(\pt_{x_2}S)^2 \pt^2_{x_1}f_k } \over
l_k^3} -
\nonumber\\
{{2ik \pt_{t_2}f_k\pt_{t_2}S + \pt^2_{x_1}f_k + ik\pt^2_{t_2}S f_k}
\over l_k^2}. \label{4th}
\end{eqnarray}
\par
The formula for the $n$-th order term has the form
$$
\stackrel{n,m}{\Psi}_{kS}=\frac1{l_k}\Big[\partial^2_{t_2}\stackrel{n-4,m}{\Psi}_{kS}+
2ik\pt_{t_2}S\partial_{t_2}\stackrel{n-2,m}{\Psi}_{kS} +
ikS_{t_2t_2}\stackrel{n-2,m}{\Psi}_{kS}- 
$$
$$
2ik\pt_{x_2}S\partial_{x_2}\stackrel{n-2,m}{\Psi}_{kS}-
ik\pt^2_{x_2}S\stackrel{n-2,m}{\Psi}_{kS} -
\partial^2_{x_1}\stackrel{n-2,m}{\Psi}_{kS}-
$$
$$
2\partial^2_{x_1x_2}\stackrel{n-3,m}{\Psi}_{kS}-
\partial^2_{x_2}\stackrel{n-4,m}{\Psi}_{kS}-
2ik\pt_{x_2}S\partial_{x_1}\stackrel{n-1,m}{\Psi}_{kS}+
$$
\bb
\gamma\!\!\!\!\!\!\!
\sum_{ \begin{array}{c}{\scriptstyle n_1+n_2+n_3=n }, \\
{\scriptstyle k_1 + k_2 +k_3 =kS }\\{\scriptstyle m_1+m_2+m_3=m}
\end{array}} \hskip-0.3cm \stackrel{n_1,m_1}{\Psi}_{k_1}
\stackrel{n_2,m_2}{\Psi}_{k_2} \stackrel{n_3,m_3}{\Psi}_{k_3} \Big]. \label{nth}
\ee
\par
The main order of the singularity corresponds  to the terms when $m=0$.
This order depends on the number of correction term.
\par
\begin{lemma} \label{lemmaAboutSingularitiesForexternalExpansion}The coefficient $\stackrel{n,m}{\Psi}_{\varphi},\ \varphi \in \Upsilon^b_{1,0}\cup \{kS \}$ has the following
behaviour
\bb
\stackrel{n,m}{\Psi}_{\varphi} = O(l_{k}^{-(n-2m-1)}),\quad l_{k} \to -0,
\label{syngord}
\ee
\end{lemma}
{\bf Proof.}
At first we prove formula (\ref{syngord}) for the phase $\varphi=kS$.  
The validity of formula (\ref{syngord}) for $n=2,3,4$ and $m=0$
directly obtains from (\ref{2th}), (\ref{3th}), (\ref{4th}). Suppose
now that this formula is valid for the term $\stackrel{n-1,0}{\Psi}_{\varphi}$.
The increase of the order of the singularity as $l_{k}\to 0$ takes place
due to differentiation with respect to $x_2, t_2$ and the nonlinear
term in formula (\ref{nth}). Differentiation of the terms in formula
(\ref{nth}) leads to formula (\ref{syngord}) for $m=0$.
\par
The validity for values $m\ge 1$ and other values of $\varphi$ can 
be obtained by direct calculations as it was shown above.  Lemma is proved.
\par
These lemma allows us to write the asymptotic representation for the coefficients
\bb
\stackrel{n,m}{\Psi}_{\varphi} =
\sum_{j=-(n-2m-1)}^{\infty}\stackrel{n,m}{\Psi}\!\!{}^j_{\varphi}\ l_{k}^j,\quad \varphi \in \Upsilon^b_{1,0}\cup \{kS \},\ l_{k}
\to -0. \label{externalAsymptoticsCloseToSingularity}
\ee
The terms $\stackrel{n,m}{\Psi}_{\varphi}$ for $\varphi \notin \Upsilon^b_{1,0}\cup \{kS \}$ have the more weaker order of the 
singularity as $l_{k} \to -0$. Equations (\ref{nth}) for these values
of $\varphi$ have the same structure but the multiplier ${{1}\over{l_j}}$
is regular as $l_{k} \to -0$.
\par
The domain of the validity as $l_{k}\to -0$ for the formal asymptotic solution
in the form (\ref{solution_before_k}) follows from the relation
$$
{\ve{\Psi^b_{n+1}}\over {\Psi^b_n}} \ll 1.
$$
It yields
$$
-l_{k} \gg \ve.
$$
The following theorem is proved.
\par
\begin{theorema}
In the domain $-l_{k}\gg \ve,\ l_{k+1} \gg \ve$ the formal asymptotic solution of
equation (\ref{we}) has form
(\ref{solution_before_k}). The coefficients of the asymptotics
 are defined either  from differential equations (\ref{nls_pre}), (\ref{lSh_pre}) or from  algebraic equations
(\ref{2th}), (\ref{3th}), (\ref{4th}), (\ref{nth}).
\end{theorema}

\section{Internal asymptotics}\label{internalAsymptotics}

This part of the paper contains the asymptotic construction of the solution for equation (\ref{we}) in the neighborhood of the curve $l_k=0$. The domain of validity of this internal  asymptotics intersects with domain of validity of expansion (\ref{solution_before_k}). These expansions are matched. Here we present asymptotic constructions for arbitrary $k$ in details. It is necessary to note this section also describes the transition from the forced oscillations of $O(\ve^2)$ to  the solution of $O(\ve)$ when $k=N$.
\par
Let us construct the internal asymptotic expansion in the domain $|l_k|\ll 1$. Denote
\bb
\l_k(x_1,t_1,\ve)={1\over \ve}l_k(\ve x_1,\ve
 t_1). \label{definitionOfTheLambda}
\ee
\par
\begin{theorema}\label{internalAsymptoticTheorem}
In the domain $|l_k|\ll 1$ the formal asymptotic solution for
equation (\ref{we}) has the form
\begin{eqnarray}
U(x,t,\ve)=\ve W_1 + \sum_{n=2}^\infty\ve^n W_n, \label{int1}
\end{eqnarray}
where
$$
W_1=\stackrel{1,0}{W}_{kS}\exp\{ik{S(t_2,x_2)/\ve^2}\}+
\sum_{m=k+1}^{N} \sum_{\pm\Phi_m}\exp\{\pm
i\Phi_m(x_2,t_2)/\ve^2\} \stackrel{1,0}{W}{}_{\pm\Phi_m}(x_1,t_1,t_2)
$$
\begin{eqnarray*}
W_n = \sum_{j=0}^{n-2}\ln^j(\ve)\stackrel{n,j}{W}_{kS}\exp\big(ik{S(t_2,x_2)\over\ve^2}\big)+ \\
\sum_{j=0}^{n-2} \ln^j(\ve)\bigg( \sum_{m=k+1}^{N}
\sum_{\pm\Phi_m}\exp\{\pm
i\Phi_m(x_2,t_2)/\ve^2\} \stackrel{n,j}{W}{}_{\pm\Phi_m}(x_1,t_1,t_2)+
\\
\sum_{\chi \in \Upsilon'_{n,j}}\exp\{i\chi(x_2,t_2)/\ve^2\}
\stackrel{n,j}{W}{}_{\chi}(x_1,t_1,t_2)\bigg), 
\end{eqnarray*}
where $\Upsilon'_{n,j}$ is the set of phase functions which is determined by 
$$
\Upsilon_{1,0}={\pm\Phi_{k+1},\pm\Phi_{k+2},\dots,\pm\Phi_{N},\pm kS};\quad 
$$
$$
\Upsilon_{2,0}={\pm\Phi_{k+1},\pm\Phi_{k+2},\dots,\pm\Phi_{N}, 
\pm S,\dots, \pm N S},
$$
\bb
\Upsilon_{n,j}=\bigcup_{\begin{array}{c} n_1+n_2+n_3=n,\\ j_1+j_2+j_3=j \end{array}}
\chi_{n_1,j_1}+\chi_{n_2,j_2}+\chi_{n_3,j_3},\,\,\,\chi_{n_p,j_p} \in \Upsilon_{n_p,j_p}.\nonumber
\ee
$$
\Upsilon'_{n,j} = \Upsilon_{n,j} \backslash \Upsilon_{1,0}.
$$
The function $\stackrel{n,j}{W}_{kS}$ is solution of the problem for equation (\ref{eqForN1-thOrderTerm}) with zero condition as $\l_k\to-\infty$.
\end{theorema}
\par
There is a difference between asymptotics (\ref{int1}) and external pre-resonance     asymptotics (\ref{solution_before_k}). Asymptotic expansion (\ref{int1}) defines the generation of the $k$-th mode by the local resonance for the leading-order term of the asymptotic expansion. The  difference is the collection of phase function. The collection of phases $\Upsilon_{1,0}$ contains $(N-k+1)$ phases but the phase $\pm\Phi_k$ does not exist in the $k$-th resonant layer. There phases $\pm k S$ are presented in the $\Upsilon_{1,0}$ and $\Upsilon_{2,0}$ collections.
\par
The proof of theorem \ref{internalAsymptoticTheorem} consists in three steps. First we derive  equations for the coefficients of the asymptotics. Second we solve the problems for the coefficients. And third we determine the domain of the validity for expansion (\ref{int1}).

\subsection{The equations for coefficients}
\par
In the domain $1 \ll \l_k \ll \ve^{-1}$ both asymptotics 
(\ref{solution_before_k}) and (\ref{int1}) are valid. This fact allows us to obtain the asymptotic representation for coefficients of (\ref{int1}). Substitute $l_k=\ve\l_k$ in formula
(\ref{externalAsymptoticsCloseToSingularity}) and expand the obtained expression with respect to powers of  $\ve$. It yields
\bb
\stackrel{n,m}{W}_{\varphi} =\sum_{j=n-2m-1}^\infty\l_k^{-j}\stackrel{n+1,m}{\Psi}{}^j_{\varphi}(x_2,t_2,x_1), \quad  \l_k \to -\infty.
\label{internalAsymptoticsCloseTo-Infinity}
\ee
\par
Let us obtain the differential equations for the coefficients of asymptotics (\ref{int1}). Substitute (\ref{int1}) in equation (\ref{we}) and collect the terms with equal powers of small parameter and exponents. It yields the equations for coefficients $\stackrel{n,j}{W}_\varphi,$ $\varphi \in \Upsilon_{n,j}.$ In particularly, the terms of the order $\ve^2$ give us the equations for the amplitudes of the leading-order terms 
\bb
2ik\pt_{t_2}S\pt_{t_1}\stackrel{1,0}{W}_{kS} - 2ik\pt_{x_2}S \pt_{x_1}
\stackrel{1,0}{W}_{kS} - \lambda_k \stackrel{1,0}{W}_{kS} = f_k,
\label{eqForLeadingOrderTerm}
\ee
\bb
2i\pt_{t_2}\Phi_j\pt_{t_1}\stackrel{1,0}{W}_{\Phi_j} - 2i\pt_{x_2}\Phi_j \pt_{x_1}
\stackrel{1,0}{W}_{\Phi_j} - l_{\Phi_j}(x_2,t_2)\stackrel{1,0}{W}_{\Phi_j} = 0,\quad j=k+1,\dots,N.
\label{eqForLeadingOrderTerm_old}
\ee
and complex conjugated equation for $\stackrel{1,0}{W}_{-\varphi},\ \varphi \in \Upsilon_{1,0}(k)$.
\par
The relation of the order $\ve^3$ in equation (\ref{we}) gives $2(N-k)+2$ equations:
\begin{eqnarray}
2ik\pt_{t_2}S\pt_{t_1}\stackrel{2,0}{W}_{kS} - 2ik\pt_{x_2}S
\pt_{x_1}\stackrel{2,0}{W}_{kS}- \lambda_k \stackrel{2,0}{W}_{kS} =
\pt^2_{x_1}\stackrel{1,0}{W}_{kS}-\pt^2_{t_1}\stackrel{1,0}{W}_{kS}  -
\nonumber
\\ 
-ik[\pt_{t_2}^2S - \pt_{x_2}^2S] \stackrel{1,0}{W}_{kS}  -2ik\pt_{t_2}S\pt_{t_2}\stackrel{1,0}{W}_{kS}
+2ik\pt_{x_2}S\pt_{x_2}\stackrel{1,0}{W}_{kS} \nonumber \\
 - 3\gamma \left( |\stackrel{1,0}{W}_{kS}|^2+ \sum_{m=k+1}^{N} |\stackrel{1,0}{W}_{\Phi_m}|^2 \right) \stackrel{1,0}{W}_{kS}.\label{eqForFirstCorrectionTerm}
\end{eqnarray}
\begin{eqnarray}
2i\pt_{t_2}\Phi_j\pt_{t_1}\stackrel{2,0}{W}_{\Phi_j} - 2i\pt_{x_2}\Phi_j
\pt_{x_1}\stackrel{2,0}{W}_{\Phi_j}- l_{\Phi_j}(x_2,t_2)\stackrel{2,0}{W}_{\Phi_j} = \nonumber
\\ 
\pt^2_{x_1}\stackrel{1,0}{W}_{\Phi_j}-\pt^2_{t_1}\stackrel{1,0}{W}_{\Phi_j}  -
-i[\pt_{t_2}^2\Phi_j - \pt_{x_2}^2\Phi_j] \stackrel{1,0}{W}_{\Phi_j}  -2i\pt_{t_2}\Phi_j\pt_{t_2}\stackrel{1,0}{W}_{\Phi_j}
+2i\pt_{x_2}S\pt_{x_2}\stackrel{1,0}{W}_{\Phi_j} \nonumber \\
- 3\gamma \left( |\stackrel{1,0}{W}_{kS}|^2+ \sum_{m=k+1}^{N} |\stackrel{1,0}{W}_{\Phi_m}|^2 \right) \stackrel{1,0}{W}_{\Phi_j}.\label{eqForFirstCorrectionTermOldPhases}
\end{eqnarray}
\par
The higher-order terms  are calculated by the same way. In particularly  $\stackrel{n,j}{W}_{\varphi}, \varphi \in \Upsilon_{1,0}$ is determined by differential equations. Here  we represent the equation for amplitude under $\varphi=kS$. The equations for other phases $\varphi \in \Upsilon_{1,0}$ can be obtained in a similar.
\bb
2ik\pt_{t_2}S\pt_{t_1}\stackrel{n,j}{W}_{kS} - 2ik\pt_{x_2}S \pt_{x_1}
\stackrel{n,j}{W}_{kS} - \lambda_k \stackrel{n,j}{W}_{kS} = \stackrel{n,j}{F}_k.
\label{eqForN1-thOrderTerm}
\ee
The right hand side of equation (\ref{eqForN1-thOrderTerm}) has the
form
$$
\stackrel{n,j}{F}_k= -2ik\pt_{t_2}S\pt_{t_2}\stackrel{n-1,j}{W}_{kS} +
2ik\pt_{x_2}S\pt_{x_2}\stackrel{n-1,j}{W}_{kS}+
k^2(\pt_{t_2}S)^2\stackrel{n-1,j}{W}_{kS} -
$$
$$
k^2(\pt_{x_2}S)^2\stackrel{n-1,j}{W}_{kS}-\pt_{t_1}^2\stackrel{n-1,j}{W}_{kS} +\pt_{x_1}^2\stackrel{n-1,j}{W}_{kS} - 
\pt_{t_2}\pt_{t_1}\stackrel{n-2,j}{W}_{kS}+
\pt_{x_2}\pt_{x_1}\stackrel{n-2,j}{W}_{kS}-  
$$
\begin{eqnarray}
\pt_{t_2}^2\stackrel{n-3,j}{W}_{kS} + 
\pt_{x_2}^2\stackrel{n-3,j}{W}_{kS} - \g\sum_{
\begin{array}{c}\small 
n_1+n_2+n_3=n+1,\\
j_1+j_2+j_3=j,\\
 \chi_1 + \chi_2 +\chi_3 =kS,\\ 
\chi_j\in \Upsilon_{n_j}(k),\,j=1,2,3 
\end{array}}
\stackrel{n_1,j_1}{W}_{\chi_1}
\stackrel{n_2,j_2}{W}_{\chi_2}
\stackrel{n_3,j_3}{W}_{\chi_3}.\label{rightSideOfEqForN1-thCorrectionTerm}
\end{eqnarray}
\par
The amplitudes $\stackrel{n,k}{W}_{\chi}$ as $\chi\notin \Upsilon_{1,0}$
are determined by algebraic equations
\bb
\left[-(\chi_{t_2})^2 + (\chi_{x_2})^2 + 1
\right]\stackrel{n,k}{W}_{\chi} = \stackrel{n,k}F_{\chi}. \label{algebraic}
\ee
Here the right hand side of the equation depends on previous
terms and their derivatives
$$
\stackrel{n,k}F_{\chi} =
-2i\chi_{t_2}\pt_{t_1}\stackrel{n-1,k}{W}_{\chi} +
2i\chi_{x_2}\pt_{x_1}\stackrel{n-1,k}{W}_{\chi} - 2i \chi_{t_2}
\pt_{t_2}\stackrel{n-2,k}{W}_{\chi} - i\left[\chi_{t_2 t_2} -
\chi_{x_2 x_2} \right]\stackrel{n-2,k}{W}_{\chi}-
$$
\bb
 \pt^2_{t_1
t_2}\stackrel{n-3,k}{W}_{\chi} - \pt^2_{t_2
t_2}\stackrel{n-4,k}{W}_{\chi} -\gamma
\sum_{N,k_2,l_1,l_2,m_1,m_2,\alpha,\beta,\delta}
\stackrel{N,k_2}{W}_{\alpha} \stackrel{l_1,l_2}{W}_{\beta}
\stackrel{m_1,m_2}{W}_{\delta},
\label{eqForNk-thCorrectionTerm}
\ee
where $N+l_1+m_1=n-4,\ k_2+l_2+m_2=k,\ \alpha + \beta + \delta =
\chi,\ \alpha\in \Upsilon_{N,k_2},\ \beta\in \Upsilon_{l_1,l_2},\
\delta \in \Upsilon_{m_1,m_2}.$

\subsection{The solvability of equations for higher-order terms}\label{SolvOfInternalEquations}
\par
In this  section we present the explicit form for the higher-order term
$\stackrel{n,j}{W}_{kS}$ and investigate the asymptotic behaviour as
 $\l_k\to\pm\infty$.

\subsubsection{Characteristic variables}

The function $\stackrel{n,j}{W}_{kS}$ satisfies equation
(\ref{eqForN1-thOrderTerm}). The solution is constructed by
characteristic method. Define the characteristic variables $\s,\xi$.
We choose a point $(x^0_1,t^0_1)$ such that
$\pt_{x_2}l_k|_{(x^0_1,t^0_1)}\not=0$ as origin and denote by $\s$
the variable along the characteristic family for equation
(\ref{eqForN1-thOrderTerm}). We suppose  $\s=0$ on the curve
$\l_k=0$. The variable $\xi$ mensurates the distance along the curve
$\l_k=0$ from the point $(x^0_1,t^0_1)$. This point $(x^0_1,t^0_1)$
corresponds to $\xi=0$. Let  the positive direction for parameter
$\xi$ coincides with the positive direction of $x_2$  in the neighborhood
of $(x^0_1,t^0_1)$.
\par
The characteristic equations for (\ref{eqForN1-thOrderTerm}) have a
form
\bb
{dt_1\over d\s}=2k\pt_{t_2}S(\ve x_1,\ve t_1),\quad {dx_1\over
d\s}=-2k\pt_{x_2}S(\ve x_1,\ve t_1).\label{eqForCharacteristics}
\ee
The initial conditions for the equations are
\bb
x_1|_{\s=0}=x^0_1,\quad
t_1|_{\s=0}=t^0_1.\label{initialConditionsForCharacteristics}
\ee
\begin{lemma}\label{lemmaAboutSolvabilityForCharateristicEq}
The Cauchy problem (\ref{eqForCharacteristics}), (\ref{initialConditionsForCharacteristics}) for characteristics has a solutions as
 $|\s|<c_1\ve^{-1},\quad c_1=const>0$.
\end{lemma}
{\bf Proof.} The Cauchy problem (\ref{eqForCharacteristics}),
(\ref{initialConditionsForCharacteristics}) is equivalent  to the
system of the integral equations
\bb
t_1=t^0_1+2\int_0^\s  k\pt_{t_2}S(\ve x_1,\ve t_1)d\z,\quad
x_1=x^0_1-2\int_0^\s  k\pt_{x_2}S(\ve x_1,\ve t_1)d\z.
\label{integralEqForCharacteristics}
\ee
Substitute $\tilde t_2=(t_1-t^0_1)\ve,\,\,\tilde
x_2=(x_1-x^0_1)\ve$. It yields
$$
\tilde t_2=2\int_0^{\ve \s}  k\pt_{t_2}S(\tilde x_2-\ve x^0_1,\tilde
t_2-\ve t^0_1)d\z,\quad \tilde x_2=-2\int_0^{\ve\s}
 k\pt_{x_2}S(\tilde x_2-\ve x^0_1,\tilde t_2-\ve t^0_1)d\z.
$$
The integrands are smooth and bounded functions on the plane
$x_2,t_2$. There exists the constant $c_1=\const>0$ such that the
integral operator is the contraction operator as $\ve|\s|<c_1$. Lemma
 \ref{lemmaAboutSolvabilityForCharateristicEq} is proved.
\par
{\bf Assumption.} We assume that the change of variables
$(x_1,t_1)\to(\s,\xi)$ is unique in the neighborhood of the curve
$\l_k=0$. This assumption means that the characteristics for
equation (\ref{eqForN1-thOrderTerm}) do not touch the curve $\l_k=0$.
It means
$$
\pt_{x_2}l_k\pt_{x_2}S-\pt_{t_2}l_k\pt_{t_2}S\not=0.
$$
\par
It is convenient to use the following asymptotic formulas for change
of variables  $(x_1,t_1)\to(\s,\xi)$.

\begin{lemma}\label{lemmaAboutAsymptoticsForCharacteristics}
In the domain $|\s|\ll\ve^{-1}$ the asymptotics  as $\ve\to0$ of the
solutions for Cauchy problem (\ref{eqForCharacteristics}),
(\ref{initialConditionsForCharacteristics})  have the form
\begin{eqnarray}
x_1(\s,\xi,\ve)-x^0_1(\xi)=-2\s k\pt_{x_2}S+2\sum_{n=1}^N
\ve^n\s^{n+1} g_n(\ve x_1,\ve t_1)+O(\ve^{N+1}\s^{N+2}),\qquad
\label{asymptoticsOf-x1}\\
t_1(\s,\xi,\ve)-t^0_1(\xi)=2\s k\pt_{t_2}S+2\sum_{n=1}^N
\ve^n\s^{n+1} h_n(\ve x_1,\ve t_1)+O(\ve^{N+1}\s^{N+2}),\qquad
\label{asymptoticsOf-t1}
\end{eqnarray}
where
$$
g_n=-k{d^n \over d \s^n}(\pt_{x_2}S)\bigg|_{\s=0},\quad  h_n=k{d^n
\over d \s^n}(\pt_{t_2}S)\bigg|_{\s=0}.
$$
\end{lemma}
\par
The lemma proves by integration by parts of equations
(\ref{integralEqForCharacteristics}).
\par
The next proposition gives us the asymptotic formula which connects
variables $\s$ and $\l_k$ as $\s,\l_k\to \pm\infty$.

\begin{lemma}\label{lemma_lambda_and_sigma}

Let be $\s\ll \ve^{-1}$, then:
$$
\l_k=\phi(\xi)\s+O(\ve\s^2),\quad \quad \phi(\xi)={d\l_k\over
d\s}\bigg|_{\s=0}\quad \s\to \infty.
$$
\end{lemma}
\par
{\bf Proof.} From formula (\ref{definitionOfTheLambda}) we obtain
the representation in  the form
$$
\l_k=\sum_{j=1}^\infty \l^{j}_k(x_1,t_1,\ve)\s^j\ve^{j-1},
$$
where
$$
\l_k^j(x_1,t_1,\ve)={1\over j!}{d^j\over
d\s^j}\l_k(x_1,t_1,\ve)|_{\s=0}.
$$
\par
It yields
$$
\l_k={d\l_k\over d\s}\big|_{\s=0} \s +O\big(\ve\s^2 {d^2 \l_k\over
d\s^2}\big).
$$
Let be
$$
\left|{d^2 l_k\over d\s^2}\right|\ge\const, \, \, \xi\in R.
$$
\par
The function $d\l_k/ d\s$ is not equal to  zero
$$
{d\l_k\over
d\s}={1\over2}\bigg(-k\pt_{x_2}\l_k\pt_{x_2}S+k\pt_{t_2}\l_k\pt_{t_2}S\bigg)\not=0.
$$
Let us suppose $d\l_k/ d\s>0$. It yields
$$
\l_k=\phi(\xi)\s+O(\ve\s^2),\quad \quad \phi(\xi)={d\l_k\over
d\s}\bigg|_{\s=0}
$$
The lemma is proved.

\subsubsection{Solutions of the equations for higher-order terms}
\par
The higher-order  terms $\stackrel{n,j}{W}_{\pm kS}$ are solutions of
equation (\ref{eqForN1-thOrderTerm}) with the given asymptotic
behaviour $\l_k\to-\infty$. Equation (\ref{eqForN1-thOrderTerm}) can
be written in terms of characteristic variables as
\bb
i{d\over d\s}\stackrel{n,j}{W}_{kS}-\l_k
\stackrel{n,j}{W}_{kS}=\stackrel{n,j}{F}_{kS}.
\label{characteristicEqForN1-thOrderTerm}
\ee
\begin{lemma}\label{lemmaAboutN1-thOrderTerm}
The solution of equation (\ref{eqForN1-thOrderTerm}) with the
asymptotic behaviour (\ref{internalAsymptoticsCloseTo-Infinity}) as
$\l_k\to-\infty$ has a form
\bb
\stackrel{n,j}{W}_{kS}=\exp(-i\int_0^\s d\zeta\l_k(x_1,t_1,\ve))
\int_{-\infty}^{\s} d\zeta\stackrel{n,j}{F}_{kS}(x_1,t_1,\ve)
\exp(-i\int_0^\z d\chi\l_k(x_1,t_1,\ve)). \label{N1-thOrderTerm}
\ee
\end{lemma}
\par
{\bf Proof.} By direct substitution we see that expression
(\ref{N1-thOrderTerm}) is the solution of
(\ref{characteristicEqForN1-thOrderTerm}). The asymptotics of this
solution as $\l_k\to-\infty$ can be obtained by integration by parts
and substitution
$$
{d\over d\s}=2k\pt_{t_2}S\pt_{t_1}-2k\pt_{x_2}S\pt_{x_1}.
$$
It yields
\bb
\stackrel{n,j}{W}_{kS}=\sum_{j=0}^\infty
\bigg({2k\pt_{t_2}S\pt_{t_1}-2k\pt_{x_2}S\pt_{x_1}\over
i\l_k}\bigg)^j\bigg[{\stackrel{n,j}{F}_k\over i\l_k}\bigg],\quad
\l_k\to-\infty. \label{asymptoticsForN1-thOrderTerm}
\ee
From formula (\ref{rightSideOfEqForN1-thCorrectionTerm}) we obtain
that formulas (\ref{asymptoticsForN1-thOrderTerm}) and
(\ref{internalAsymptoticsCloseTo-Infinity}) are equivalent. The lemma
is proved.

\subsection{Asymptotics as $\l_k\to\infty$ and domain of validity of
the internal asymptotics}
\par
The domain of validity of the internal expansion is determined by the
asymptotics of higher-order terms. In this section we show that the
$n-$th order term of the asymptotic solution increases as
$\l_k^{n-1}$ when $\l_k\to\infty$. This increase of higher-order
terms
 allows us to determine the domain of validity for
internal asymptotics (\ref{int1}) as $\l_k\to\infty$.
\par

\subsubsection{Asymptotics of higher-order terms} \label{asymptoticsOfHOT}
\par
This section contains two propositions concerning asymptotic
behaviour as $\l_k\to\infty$ for  higher-order terms in (\ref{int1}).
The first lemma describes the asymptotic behaviour of higher-order
terms as $\l_k\to\infty$ and the second one contains a result about
asymptotics of the phase function.
\par
\begin{lemma}\label{lemmaAboutAsymptoticsAsLambdaToInfonity}
The asymptotic behaviour of $\stackrel{n,j}{W}_{kS}$ as
$1\ll\l_k\ll\ve^{-1}$ has a form
\begin{eqnarray}
\stackrel{n,j}{W}_{kS}=\sum_{j=0}^{n-1}\sum_{m=0}^{j-1}\bigg(\l_k^j\ln^{m}|\l_k|\stackrel{n,j}{W}{}\!^{(j,m)}_k(\xi)\bigg)
\exp(-i\int_0^\s d\z \l_k(x_1,t_1,\ve))\,+\nonumber\\
+
\sum_{m=0}^{\infty}\bigg({2k\pt_{t_2}S\pt_{t_1}-2k\pt_{x_2}S\pt_{x_1}\over
i\l_k} \bigg)^m\bigg[{\stackrel{m,j}{F}_{kS}\over i\l_k}\bigg].
\label{asymptoticsForN1-thOrderTermAsPlusInfinity}
\end{eqnarray}
\end{lemma}
\par
{\bf Proof.} Let us calculate the asymptotics of the leading-order
term
$$
\stackrel{1,0}{W}_{kS}=\exp(-i\int_0^\s d\z
\l_k(x_1,t_1,\ve))\int_{-\infty}^\z d\z f_k(x_1)\exp(i\int_0^\s d\chi
\l_k(x_1,t_1,\ve))=
$$
$$
\exp(-i\int_0^\s d\z \l_k(x_1,t_1,\ve))\int_{-\infty}^\infty d\z
f_k(x_1)\exp(i\int_0^\z d\chi \l_k(x_1,t_1,\ve)) -
$$
$$
\exp(-i\int_0^\s d\z \l_k(x_1,t_1,\ve))\int_{-\s}^\infty d\z
f_k(x_1)\exp(i\int_0^\z d\chi \l_k(x_1,t_1,\ve)).
$$
Further by integration by parts of the last term we obtain formula
(\ref{asymptoticsForN1-thOrderTermAsPlusInfinity}) as $n=1$, where
$$
\stackrel{1,0}{W}{}\!^{(0,0)}_{kS}(\xi)=\int_{-\infty}^{\infty} d\s
f_k(x_1)\exp(i\int_0^\s d\chi \l_k(x_1,t_1,\ve)),
$$
$$
\stackrel{1,0}{F}_{kS}=f_k(x_1).
$$
\par
To calculate the asymptotics of $\stackrel{2,0}{W}_{kS}$ in formula
(\ref{N1-thOrderTerm}) we use the asymptotics with respect to
$\s$ of the leading-order term. Integral (\ref{N1-thOrderTerm})
contains the term with linear increase with respect to $\s$ when
$n=2$. We eliminate this growing part from integral explicitly.
The residuary integral converges as $\s\to\infty$. It can be
calculated in the same manner as it was calculated for
$\stackrel{1,0}{W}_{kS}$. It yields formula
(\ref{asymptoticsForN1-thOrderTermAsPlusInfinity}) as $n=2$, where
$$
\stackrel{2,0}{W}{}\!^{(1,0)}_{kS}(\xi)=\stackrel{1,0}{W}{}\!^{(0,0)}_{kS}(\xi).
$$
The same direct calculations are realized for the $n-$th order
 term. The lemma is proved.
\par
To complete the proof of theorem \ref{internalAsymptoticTheorem} we
need to obtain the domain of validity of asymptotics (\ref{int1}).
The formal series (\ref{int1}) is asymptotic when
$$
{\ve W_{n+1}\over W_{n}}\ll1,
\quad\ve\to0.
$$
Lemma \ref{lemmaAboutAsymptoticsAsLambdaToInfonity} gives
$\l_k\ll\ve^{-1}.$ After substitution $\l_k=\ve l_k$ we obtain
$l_k\ll 1.$ Theorem \ref{internalAsymptoticTheorem} is proved.

\subsubsection{Asymptotics of the phase function as $\l_k\to\infty$}
\label{internalAsymptoticsAsLambdaToInfinity}

To obtain the asymptotics as $\l_k\to\infty$ we need to derive the
asymptotics of the phase function in formula
(\ref{asymptoticsForN1-thOrderTermAsPlusInfinity}).
\par
\begin{lemma}\label{lemmaAboutAsymptoticsOfPhaseAsLambdaToInfinity}
As $\l_k\to\infty$:
\bb\label{asymptoticsOfPhaseAsLambdaToInfinity}
\int_0^\s d\xi \l_k={S\over\ve^2} \,+\,
{1\over\ve}(\pt_{x_2}S(x_1-x_1^0)+\pt_{t_2}S(t_1-t_1^0))+\,
O(\ve\l_k^3).
\ee
\end{lemma}
\par
{\bf Proof.} Substitute the asymptotics of $\l_k$ from lemma
\ref{lemmaAboutAsymptoticsAsLambdaToInfonity}. Calculate the
asymptotics of the integral in formula
(\ref{asymptoticsOfPhaseAsLambdaToInfinity})
$$
\int_0^\s d\z\l_k(x_1,t_1,\ve)=\int_0^\s
{d\z\over2}\bigg[(-k\pt_{x_2}l_k\pt_{x_2}S+k\pt_{t_2}l_k\pt_{t_2}S)\z\,+\,
O(\ve\z^2)\bigg]=
$$
$$
(-k\pt_{x_2}l_k\pt_{x_2}S+k\pt_{t_2}l_k\pt_{t_2}S){\s^2\over4}
+O(\ve\s^3).
$$
The asymptotics of the phase function $kS(x_2,t_2)$ in the
neighborhood of the curve $l_1=0$ is represented by a segment of the
Taylor series. It yields
$$
{kS\over\ve^2}={1\over\ve}(k\pt_{x_2}S(x_1-x_1^0)+k\pt_{t_2}S(t_1-t_1^0))+
$$
$$
{1\over2}
(kS_{x_2x_2}(x_1-x_1^0)^2+2kS_{x_2t_2}(x_1-x_1^0)(t_1-t_1^0)+
kS_{t_2t_2}(t_1-t_1^0)^2) +
$$
$$
O(\ve(|t_1-t_1^0|+|t_1-t_1^0|)^3).
$$
Substitute instead of $(x_1-x_1^0)$ and  $(t_1-t_1^0)$ their
asymptotic behaviour with respect to $\ve$ from lemma
\ref{lemmaAboutAsymptoticsForCharacteristics}. This substitution and
result of lemma \ref{lemma_lambda_and_sigma} complete the proof of
lemma \ref{lemmaAboutAsymptoticsOfPhaseAsLambdaToInfinity}.
\par
The asymptotics as $\l_k\to-\infty$ contains fast oscillating terms
with phase functions $\Phi_{k+1},\dots,\Phi_{N}$, $mS, m\in Z,\,\,m\not=k$. The leading-order term of the
asymptotics as $\l_k\to\infty$ contains the oscillations with an
additional phase function. We obtain this result from lemma
\ref{lemmaAboutAsymptoticsAsLambdaToInfonity}. Denote this new phase
function by $\Phi_{k}(x_2,t_2)/\ve^2$. The asymptotics of this function
is obtained in lemma
\ref{lemmaAboutAsymptoticsOfPhaseAsLambdaToInfinity}. The
nonlinearity and additional phase function lead to more complicated
structure of the phase set for higher-order terms of the asymptotics
as $\l_k\to\infty$.
\par
\begin{lemma}\label{lemmaAboutPhasesInAsymptoticsAsLambdaToInfinity}
The phase set $\Upsilon_{n,j}$ for the $n-$th order term of the
asymptotics as $\l_k\to\infty$ is determined by formula
$$
\Upsilon_{1,0}=\{\pm\Phi_{k},\dots,\pm\Phi_{N}\};\quad \Upsilon_{2,0}=\{\pm\Phi_{k},\dots,\pm\Phi_{N},\pm S,\dots, \pm
N S\},
$$
$$
\Upsilon_{n,j}=\cup 
\chi_{n_1,j_1}+\chi_{n_2,j_2}+\chi_{n_3,j_3},\quad \chi_{n_m,j_m} \in \Upsilon_{n_m,j_m},
$$
where $n_1+n_2+n_3=n, j_1+j_2+j_3=j$.
\end{lemma}
The proof of this lemma follows from the asymptotic formula for
$n-$th order term. Representation (\ref{int1}), formula
(\ref{asymptoticsForN1-thOrderTermAsPlusInfinity}) and lemma
\ref{lemmaAboutAsymptoticsAsLambdaToInfonity} allow us to construct
the asymptotics as $\l\to\infty$ of the internal expansion in an
explicit form
\begin{eqnarray}
U=\sum_{n=1}^N\ve^n \sum_{p=0}^{n-2} \ln^p(\ve) \bigg(\sum_{j=0}^{n-1}\sum_{m=0}^{n-2}\l^j\ln^{m}|\l|\stackrel{n,p}{W}{}\!^{(j,m)}_k(\xi)\bigg)
\times\nonumber\\
\exp\bigg[-i\bigg({k\over\ve}(\pt_{x_2}S(x_1-x_1^0)+\pt_{t_2}S(t_1-t_1^0))+\,
O(\ve\l^3)\bigg)\bigg]\, +\,
\nonumber\\
\sum_{n=1}^N\ve^n\sum_{p=0}^{n-2}\ln^p(\ve)\bigg(
\sum_{j=0}^{\infty}\bigg({2k\pt_{t_2}S\pt_{t_1}-2k\pt_{x_2}S\pt_{x_1}\over
i\l} \bigg)^j\bigg[{\stackrel{n,p}{F}_{kS}\over
i\l}\bigg]\bigg)\exp\left\{ik{{S(t_2,x_2)}\over {\ve^2}}\right\}+
\nonumber\\
\sum_{n=2}^N\ve^n\sum_{p=0}^{n-2}\ln^p(\ve)\bigg(\sum_{\varphi\in\Upsilon_{n,p}\backslash {kS}
}\stackrel{n,p}{W}_{\varphi}\exp\left\{i{{\varphi(t_2,x_2)}\over
{\ve^2}}\right\}\bigg)+c.c.
\label{asymptoticOfTheSolutionAsLambdaToInfinity}
\end{eqnarray}
This representation and formula (\ref{eqForNk-thCorrectionTerm})
complete the proof of the lemma.

\section{Post-resonance expansion}\label{Post_resonance_expansion}
\par
In this section we show how to connect the external asymptotic solution after the passage through the resonance with the solution in the neighborhood of the curve $l_k=0$.
\par
There is a difference between these solution. The  leading-order term of the solution in this section includes an additional mode. This new mode relates to the phase $\Phi_k$ which was generated by the passage through  the resonance near the curve $l_{k}=0$. It gives 
\bb
\Upsilon^{p}_{1,0} = \Upsilon^{b}_{1,0} \cup \{\pm \Phi_k\} \label{changingPhaseSet}
\ee
\par
This section consists in two parts. The first
part contains the construction of the formal asymptotic solution. This solution is similar to the solution from section \ref{externalAsymptotics1}.
 Asymptotic behaviour for higher-order terms as $l_k \to 0$ follows
from section \ref{internalAsymptoticsAsLambdaToInfinity}. In the
second part of this section we  determine  the domain of validity
for this external asymptotics near resonance curve $l_{k}(x_2,t_2)=0$.
The matching method gives us the initial conditions for the coefficients of the asymptotics.
\par

\subsection{Structure of the second external asymptotics}
\label{second_external_asymptotics}
\par
Let us construct the formal asymptotic solution of the following form  with the changing of the phase set according to (\ref{changingPhaseSet}).  
\bb
U(x,t,\ve) = \ve \Psi^{p}_{1}+ \sum_{n=2}^\infty\ve^n \Psi^{p}_{n},\label{solution_after_k}
\ee
$$
\Psi^{p}_{1}=\sum_{m=k+1}^{N} \sum_{\pm\Phi_m}\exp\{\pm
i\Phi_m(x_2,t_2)/\ve^2\} \stackrel{1,0}{\Psi}{}_{\pm\Phi_m}(x_1,t_1,t_2)
$$
\begin{eqnarray*}
\Psi^{p}_{n} = \sum_{j=0}^{n-2} \ln^j(\ve)\sum_{m=k+1}^{N}
\bigg(\sum_{\pm\Phi_m}\exp\{\pm
i\Phi_m(x_2,t_2)/\ve^2\} \stackrel{n,j}{\Psi}{}_{\pm\Phi_m}(x_1,t_1,t_2)+ \\
 \sum_{\chi \in
\Upsilon^{p}{}'_{n,j}}\exp\{i\chi(x_2,t_2)/\ve^2\}
\stackrel{n,j}{\Psi}{}_{\chi}(x_1,t_1,t_2)\bigg),
\end{eqnarray*}
where $\Upsilon^{p}{}'_{n,j}$ is a set of phase functions which is determined by 
$$
\Upsilon^p_{1,0}=\{\pm\Phi_{k},\pm\Phi_{k+1},\dots,\pm\Phi_{N}\};\quad \Upsilon^p_{2,0}= \Upsilon^p_{1,0} \cup \{\pm S,\dots, \pm N S\},
$$
\bb
\Upsilon^p_{n,j}=\bigcup_{\begin{array}{c} n_1+n_2+n_3=n,\\ j_1+j_2+j_3=j \end{array}}
\chi_{n_1,j_1}+\chi_{n_2,j_2}+\chi_{n_3,j_3},\,\,\,\chi_{n_p,j_p} \in \Upsilon^p_{n_p,j_p}.\nonumber
\ee
$$
\Upsilon^{p}{}'_{n,j} = \Upsilon^p_{n,j} \backslash \{\pm\Phi_{k},\pm\Phi_{k+1},\dots,\pm\Phi_{N}\}.
$$
Substitute this asymptotic solution in
original equation and collect the terms of the same order with
respect to $\ve$.  After collecting the terms with the same phase functions
we obtain the recurrent system of equations for the coefficients.
\par
The terms of
the order $\ve^1$ give us the equation (\ref{equations_for_phases_pre}) for the phase
function of eigen oscillations. The initial data is determined by
matching condition (\ref{inialDataForPhases}) with the changing of indexes from $(k+1)$ to $k$.
\par
The difference between this asymptotic constructions and constructions from section \ref{externalAsymptotics1} consists in changing of the number of terms of the order of $\ve$.
\par
The amplitudes $\stackrel{1,0}{\Psi}{}_{\Phi_j}$ are determined from 
\begin{eqnarray}
2i\pt_{t_2}\Phi_j\pt_{t_2}\stackrel{1,0}{\Psi}{}_{\Phi_j} +
[(\pt_{t_1}\xi_j)^2-(\pt_{x_1}\xi_j)^2]\pt^2_{\xi_j\xi_j}\stackrel{1,0}{\Psi}{}_{\Phi_j}
+  \nonumber\\
+i[\pt_{t_2}^2\Phi_j -
\pt_{x_2}^2\Phi_j]\stackrel{1,0}{\Psi}{}_{\Phi_j} + \gamma
|\stackrel{1,0}{\Psi}{}_{\Phi_j}|^2 \stackrel{1,0}{\Psi}{}_{\Phi_j}
=0,\quad j=k,\dots,N \label{system_nls_post}
\end{eqnarray}
\par
The higher order terms of (\ref{solution_after_k}) satisfy  equation (\ref{lSh_pre}). Here we present the following lemma which allows to match solution (\ref{solution_after_k}) and solution (\ref{int1}).
\par
\begin{lemma}\label{lemmaAboutAsymptoticsForLS}
The asymptotics as $l_k\to 0$ of the solution of equation
(\ref{lSh_pre}) has the  form
\bb
\stackrel{n,p}{\Psi}{}_{\Phi_k}(x_1,t_1,t_2) =
\sum_{j=-(n-2p-2)}^{1}\sum_{m=0}^{-(j-1)}
\stackrel{n,p}{\Psi}{}_{\Phi_k}^{j,m}(x_1,t_1)\ l_k^j(\ln l_k)^m +
O(1), \quad l_k\to 0. \label{asymptotics_for ls}
\ee
\end{lemma}
{\bf Proof.} Determine the order of the singularity of the right
hand side of the equation as $l_k\to 0$. First consider  equation
(\ref{lSh_pre}) for $n=3, p=0$. The solution of this equation gives us
the coefficient $\stackrel{3,0}{\Psi}{}_{\Phi_k}$. The nonlinearity
contains the term  $|\stackrel{2,0}{\Psi}_{kS}|^2
\stackrel{1,0}{\Psi}_{\Phi_k}$. The function
$\stackrel{2,0}{\Psi}_{kS}$ has the singularity of the order
$l_k^{-1}$ as $l_k\to 0$. It determines the order of singularity for
right hand side $l_k^{-2}$. We construct the asymptotics of
$\stackrel{3,0}{\Psi}{}_{\chi}$ in the form
\bb
\stackrel{3,0}{\Psi}{}_{\Phi_k} =
\stackrel{3,0}{\Psi}{}_{\Phi_k}^{-1,0} l_k^{-1} +
\stackrel{3,0}{\Psi}{}_{\Phi_k}^{0,1} \ln(l_k) +
\stackrel{3,0}{\Psi}{}_{\Phi_k}^{1,1} l_k \ln(l_k) + \widehat{
\stackrel{3,0}{\Psi}}{}_{\Phi_k}, \label{asym_n3}
\ee
Substitute (\ref{asym_n3}) in equation for $n=3$. It leads to
recurrent system of equations for coefficients
$\stackrel{3,0}{\Psi}{}_{\Phi_k}^{(j,p)}$
$$
-2i
\pt_{t_2}\Phi_k\pt_{t_2}l_k\stackrel{3,0}{\Psi}{}_{\Phi_k}^{(-1,0)}
= - \stackrel{1,0}{\Psi}_{\Phi_k} |\stackrel{2,0}{\Psi}{}_{S}|^2
l_k^2,
$$
$$
2i\pt_{t_2}\Phi_1\pt_{t_2}l_k\stackrel{3,0}{\Psi}{}_{\Phi_k}^{(0,1)}=
L[\stackrel{3,0}{\Psi}{}_{\Phi_k}^{(-1,0)}],
$$
$$
2i\pt_{t_2}\Phi_k\pt_{t_2}l_k\stackrel{3,0}{\Psi}{}_{\Phi_k}^{(1,1)}=
L[\stackrel{3,0}{\Psi}{}_{\Phi_k}^{(0,1)}].
$$
Here we denote the linear operator by
$$
L[\Psi]=2i\pt_{t_2}\Phi_k\pt_{t_2}\Psi+\pt_\xi^2\Psi+i[\pt_{t_2}^2\Phi_k-\pt_{x_2}^2\Phi_k]\Psi+
\gamma\big(2|\stackrel{1,0}{\Psi}_{\Phi_k}|^2\Psi+
(\stackrel{1,0}{\Psi}_{\Phi_k})^2\Psi^*\big).
$$
\par
The regular part $\widehat{\stackrel{3,0}{\Psi}{}_{\Phi_k}}$ of the
asymptotics satisfies the nonhomogeneous linear Schrodinger
equation. The right hand side of the equation is smooth
$$
L[\widehat{
\stackrel{3,0}{\Psi}}{}_{\Phi_k}]=-l_k\ln|l_k|L[\stackrel{3,0}{\Psi}{}^{(1,1)}_{\Phi_k}]-
2i\pt_{t_2}\Phi_k\pt_{t_2}l_k\stackrel{3,0}{\Psi}{}^{(1,1)}_{\Phi_k}.
$$
The initial condition for the regular part of the asymptotics is
determined below by matching with the internal asymptotic expansion.
\par
The structure of the terms $\stackrel{n,p}{\Psi}_{\pm\Phi_k}$ for
$n>3$ has a similar form. The right hand side of equation
(\ref{lSh_pre}) depends on junior terms. These singularities can be
eliminate
$$
\stackrel{n,p}
F_{\Phi_k}=\sum_{j=0}^{-(n-2)}\sum_{m=0}^{-j+1}l_k^j\ln^m|l_k|
\stackrel{n,p}{f_k}{}^{(j,m)}_{\Phi_1}+\widehat{
\stackrel{n,p}{F_k}}_{\Phi_1}.
$$
The coefficients $\stackrel{n,p}{f_k}{}^{(j,m)}_{\Phi_k}$  do not
contain singularities as $l_k\to 0$. These coefficients are easy
calculated.
\par
The direct substitution of (\ref{asymptotics_for ls}) in  equation
and collecting the terms with the same order of $l_k$ complete the
proof of lemma \ref{lemmaAboutAsymptoticsForLS}.

\subsection{The domain of validity of the second external asymptotics
as $l_k \to +0$ and matching procedure}
\par
The domain of validity of the second external asymptotics is
determined by
$$
{\ve\Psi_{n+1}^p\over\Psi_{n}^p}\ll1.
$$
Formulas (\ref{solution_after_k}) and  (\ref{asymptotics_for ls}) give the
condition
$$
l_k \gg \ve.\label{validityOfSecondExtAs}
$$
\par
The domain $|l_k| \ll 1$ of validity of the internal asymptotics and
domain of validity of the second external asymptotics are
intersected. This fact allows to complete  the construction of the
second external asymptotics by matching method \cite{Il'in}.  The
structure of singular parts of the internal asymptotics as  $\l_k\to
+\infty$ and external asymptotics as $l_k\to 0$ are equivalent. The
coefficients are coincided due to our constructions. The matching of
regular parts of these asymptotics takes place due to
$$
\stackrel{n,j}{\Psi}_{\Phi_k}|_{l_k=0}=\stackrel{n,j}{W}{}_{kS}^{(0,0)}(\xi).
$$
The function  $\stackrel{n,j}{W}{}_{kS}^{(0,0)}(\xi)$ is determined in lemma
\ref{lemmaAboutAsymptoticsAsLambdaToInfonity}.
\par
In particular, the initial condition for the amplitude $\stackrel{1,0}{\Psi}{}_{\Phi_k}$ corresponding to the new generated phase $\Phi_k$  has a form
$$
\stackrel{1,0}{\Psi}{}_{\Phi_k}|_{l_k=0}= \int_{-\infty}^{\infty} d\s
f_k(x_1)\exp(i\int_0^\s d\chi \l_k(x_1,t_1,\ve)).
$$
The initial data for the others amplitudes $\stackrel{1,0}{\Psi}{}_{\Phi_j},\quad j=k+1,\dots,N$ of the leading-order term of (\ref{solution_after_k}) are represented by values of these amplitudes on the curve $l_k=0$.
\par
The soliton theory for nonlinear Schrodinger equation leads us to
the fact that the function $\stackrel{1,0}{\Psi}{}_{\Phi_j}$ contains
the solitary waves when $f_j(x_1)$ is sufficiently large.
\par

\section{Resonances in the higher order terms of
the asymptotics.} \label{ResonancesInHigherOrders}
\par
In this section we discuss the passage through the resonances 
 of higher order terms of the asymptotic solution only. Such passage does not change the leading-order term  and leads to changing of the solution in higher order terms only.
\par
Equations (\ref{algebraic}) are solvable while the multiplier
$$
l[\chi]=\left[-(\chi_{t_2})^2 + (\chi_{x_2})^2 + 1 \right]\not=0.
$$
Let us denote a particular value of $\chi$  such that $l[\chi]=0$ by $\chi_r$.
In the neighborhood of the curve $l[\chi_r]=0, \chi_r\in \Upsilon_{n,k}$
the coefficients of the asymptotic expansion has a form
\bb
\stackrel{n,k}{\Psi}_{\chi_r} = {\cal O}(l^{-1}[{\chi_r}]), \quad l[{\chi_r}]\to
0, \label{ordersin1}
\ee
\bb
\stackrel{n+m,k}{\Psi}_{\chi_r} = {\cal O}(l^{-(m+1)}[{\chi_r}]), \quad
l[{\chi_r}]\to 0 \label{ordersin1}
\ee
These formulas give the domain of validity of (\ref{solution_after_k})
\bb
l[{\chi_r}]\gg \ve. \label{domain_of_validity_lpsi}
\ee
In the neighborhood of the curve $l[{\chi_r}]=0$ we use a new scaled
variable $\lambda_{\chi_r}=l_{\chi_r}/\ve$. The formal asymptotic solution
is constructed in the form
\bb
U(x,t,\ve)=U_{n-2}(x,t,\ve)+U_{res}(x,t,\ve). \label{inexlpsi}
\ee
The asymptotic solution has two parts. The first one does not
depend on the scaled variable $\lambda_{\chi_r}$. But the second part
of the solution depends on $\lambda_{\chi_r}$, because the resonance
on the curve $\lambda_{\chi_r}=0$ appears only in higher order
terms of the asymptotics. The terms $U_{n-2}(x,t,\ve)$ and 
$U_{res}(x,t,\ve)$ have a form
\begin{eqnarray}
U_{n-2}(x,t,\ve)=\sum_{m=1}^{n-2}\ve^m \sum_{k=0}^{m-2} \ln^k(\ve)
\bigg(\sum_{\pm\Phi}\exp\{\pm
i\Phi(x_2,t_2)/\ve^2\} \stackrel{m,k}{\Psi}{}_{\pm\Phi}(x_1,t_1,t_2)+\nonumber\\
 \sum_{{\chi_r} \in
\Upsilon'_{m,k}}\exp\{i{\chi_r}(x_2,t_2)/\ve^2\}
\stackrel{m,k}{\Psi}{}_{{\chi_r}}(x_1,t_1,t_2)\bigg). \label{unm2}
\end{eqnarray}
\begin{eqnarray}
U_{res}(x,t,\ve)=\sum_{m=n-1}^\infty\ve^m \sum_{k=0}^{m-2}
\ln^k(\ve) \bigg(\sum_{\pm\Phi}\exp\{\pm
i\Phi(x_2,t_2)/\ve^2\} \stackrel{m,k}{\Psi}{}_{\pm\Phi}(\lambda_{\chi_r},x_1,t_1,t_2)+\nonumber\\
 \sum_{{\chi_r} \in
\Upsilon'_{m,k}}\exp\{i{\chi_r}(x_2,t_2)/\ve^2\}
\stackrel{m,k}{\Psi}{}_{{\chi_r}}(\lambda_{\chi_r},x_1,t_1,t_2)\bigg).
\label{unm1}
\end{eqnarray}
Substitution of (\ref{inexlpsi}) into (\ref{we}) gives the
recurrent sequence of the problems for the coefficients of
(\ref{inexlpsi}). Note that the coefficients
$\stackrel{m,k}{\Psi}{}_{\pm\Phi}$ and
$\stackrel{m,k}{\Psi}{}_{{\chi_r}}$ for $m\le (n-1)$ are determined
from the standard problems as shown in section
\ref{Post_resonance_expansion}.
\par
The coefficients $\stackrel{m,k}{\Psi}{}_{\pm\Phi}$ and
$\stackrel{m,k}{\Psi}{}_{{\chi_r}}$ for $m\ge (n-1)$ are determined
as well as the internal expansion from section
\ref{internalAsymptotics}. Here we present the standard problems
for the coefficients of the asymptotics without detailed
derivation.
The higher-order terms  are calculated by the same way. In
particularly, the  terms in the case lower index is equal to ${\chi_r}$ are
determined by differential equations.
\bb
2i\pt_{t_2}{\chi_r}\pt_{t_1}\stackrel{m,k}{\Psi}_{\chi_r} - 2i\pt_{x_2}{\chi_r} \pt_{x_1}
\stackrel{m,k}{\Psi}_{\chi_r} - \lambda_{\chi_r} \stackrel{m,k}{\Psi}_{\chi_r} = \stackrel{m,k}{F}_{\chi_r}.
\label{eqForInternalCorrectionInTheResonanceLayer}
\ee
The right hand side $\stackrel{m,k}{F}_{\chi_r}$ of equation (\ref{eqForInternalCorrectionInTheResonanceLayer}) has 
form  (\ref{rightHandSideForAlgebraic}) with the changing of the index from $\chi$ to $\chi_r$.
\par
The terms in the case the lower index is not equal to
${\chi_r}$ are determined by algebraic equations (\ref{algebraic}).
\par
The analysis of equation (\ref{eqForInternalCorrectionInTheResonanceLayer}) is realized in section \ref{SolvOfInternalEquations}. The passage through this resonance layer leads to a new phase function appearance as was shown in 
 \ref{internalAsymptoticsAsLambdaToInfinity}. We denote this new phase function by $\varphi_r$. 
\par
In the domain after passage through resonance layer the amplitude under this phase function is determined by
\begin{eqnarray}
2i\pt_{t_2}\varphi_r\pt_{t_2}\stackrel{n-1,k}{\Psi}{}_{\varphi_r} +
[(\pt_{t_1}\xi_r)^2-(\pt_{x_1}\xi_r)^2]\pt^2_{\xi_r\xi_r}\stackrel{n-1,k}{\Psi}{}_{\varphi_r}
+  \nonumber\\
+i[\pt_{t_2}^2\varphi_r -
\pt_{x_2}^2\varphi_r]\stackrel{n-1,k}{\Psi}{}_{\varphi_r} + \gamma
|\stackrel{1,0}{\Psi}{}_{\Phi_1}|^2 \stackrel{n-1,k}{\Psi}{}_{\varphi_r}
=0, \label{LsForHighOrderCorrectionTerms}
\end{eqnarray}
here the variable $\xi_r$ is determined by 
$$
{{dx_1}\over{d\xi_r}}=\pt_{t_2}\chi_r,\quad
{{dt_1}\over{d\xi_r}}=\pt_{x_2}\chi_r.
$$

After the passage of the  resonance layer the set $\Upsilon_{n-1,k}$ is changed by the following rule
$$
\Upsilon_{n-1,k} \rightarrow  \Upsilon_{n-1,k} \bigcup \chi_r.
$$
It leads to changing of the set of phases for higher order correction terms by the ordinary way. 
\par
{\bf Acknowledgments.} We are grateful to  I.V. Barashenkov,
L.A. Kalyakin and B.I.Suleimanov for helpful comments
and for help in improving of the mathematical presentation the
results.

\end{document}